\documentclass[%
reprint,
superscriptaddress,
amsmath,amssymb,
aps,
prl,
]{revtex4-1}
\usepackage[utf8]{inputenc}
\usepackage[pdftex]{graphicx}
\usepackage{dcolumn}
\usepackage{mathrsfs}
\usepackage{bm}
\usepackage[dvipdfmx]{color}
\usepackage{txfonts}
\usepackage{braket}
\usepackage{amsmath}
\usepackage{mathtools}

\begin{document}

\title{Elemental Dilution Effect on the Elastic Response \\due to a Quadrupolar Kondo Effect of the Non-Kramers System $\rm{Y_{1-\it{x}}Pr_{\it{x}}Ir_{2}Zn_{20}}$}
\author{R. Hibino}
 \affiliation{Department of Physics, Hokkaido University, Sapporo 060-0810, Japan}
\author{T. Yanagisawa}
 \affiliation{Department of Physics, Hokkaido University, Sapporo 060-0810, Japan}
 \author{Y. Mikami}
 \affiliation{Department of Physics, Hokkaido University, Sapporo 060-0810, Japan}
\author{H. Hidaka}
 \affiliation{Department of Physics, Hokkaido University, Sapporo 060-0810, Japan}
\author{H. Amitsuka}
 \affiliation{Department of Physics, Hokkaido University, Sapporo 060-0810, Japan}
\author{S. Zherlitsyn}
 \affiliation{Hochfeld-Magnetlabor Dresden (HLD-EMFL) and W\"{u}rzburg-Dresden Cluster of Excellence ct.qmat, Helmholtz-Zentrum Dresden-Rossendorf, 01328 Dresden, Germany}
\author{J. Wosnitza}
 \affiliation{Hochfeld-Magnetlabor Dresden (HLD-EMFL) and W\"{u}rzburg-Dresden Cluster of Excellence ct.qmat, Helmholtz-Zentrum Dresden-Rossendorf, 01328 Dresden, Germany}
 \affiliation{Institut fur Festk\"{o}rper- und Materialphysik, TU Dresden, 01062 Dresden, Germany}
\author{Y. Yamane}
\affiliation{Graduate School of Science, University of Hyogo, Kamigori 678-1297, Japan}
\author{T. Onimaru}
\affiliation{Graduate School of Advanced Sciences of Matter, Hiroshima University, Higashi-Hiroshima 739-8530, Japan}

\date{\today}
             
\begin{abstract}
We measured the elastic constants $\left(C_{11}-C_{12}\right)/2$ and $C_{44}$ of the non-Kramers system $\rm{Y_{0.63}Pr_{0.37}Ir_{2}Zn_{20}}$ (Pr-37\% system) by means of ultrasound to check how the single-site quadrupolar Kondo effect is modified by increasing the Pr concentration. 
The Curie-like softening of $\left(C_{11}-C_{12}\right)/2$ of the present Pr-37\% system on cooling from 5 to 1 K can be reproduced by a multipolar susceptibility calculation based on the non-Kramers $\Gamma_3$ doublet crystalline-electric-field ground state. Further, on cooling below $0.15\ \rm{K}$, a temperature dependence proportional to $\sqrt{T}$ was observed in $\left(C_{11}-C_{12}\right)/2$. This behavior rather corresponds to the theoretical prediction of the quadrupolar Kondo ``lattice'' model, unlike that of the Pr-3.4\% system, which shows a logarithmic temperature dependence based on the ``single-site'' quadrupolar Kondo theory.
In addition, we discuss the possibility to form a vibronic state by the coupling between the low-energy phonons and the electric quadrupoles of the non-Kramers doublet in the Pr-37\% system, since we found a low-energy ultrasonic dispersion in the temperature range between 0.15 and 1 K.\end{abstract}

\maketitle 

\section{Introduction}
\vspace{-2mm}
 In metallic compounds containing magnetic atoms, a variety of physical properties emerges due to the interaction between the localized moments and the conduction electrons. The interaction between localized ``magnetic'' moments and the spins of conduction electrons is the origin of the (magnetic) Kondo effect \cite{Kondo,Yoshida}. On the other hand, the quadrupolar Kondo effect (QKE), which originates from ``electric'' quadrupolar degrees of freedom, may also be relevant \cite{Cox.1}. Many experimental investigations have been carried out on uranium-based compounds to search for experimental evidence of the QKE \cite{Cox.1,Seaman,Aliev,Amitsuka}. However, the experimental validation of this theory in uranium compounds has been challenging, because of the duality of the 5$f$ electrons, i.e., their partially localized and itinerant character, and the associated uncertainty in the valence of the uranium ion. In contrast, cubic Pr systems with a non-Kramers doublet crystalline-electric-field (CEF) ground state, in which the $\Gamma_3$ electric quadrupole can be active, have been studied as another candidate material class to exhibit the QKE \cite{Yatskar, Bucher, Onimaru.1, Tanida}.
 
Recently, intriguing phenomena caused by the electric quadrupolar degrees of freedom have been found in the $\rm{Pr\it{T}_{\rm{2}}\it{X}_{\rm{20}}}$ ($T$ = Ir, Rh, V, and Ti; $X$ = Al, Zn, and Cd) systems, which crystallize in the cubic $\rm{CeCr_{2}Al_{20}}$-type structure ($Fd\bar{3}m$, $O_{\rm h}^{\rm 7}$, No. 227) as shown in Fig. 1 \cite{Nasch}. The title compound $\rm{PrIr_2Zn_{20}}$ is one of them, showing an antiferroquadrupolar (AFQ) and superconducting transitions at $T_{\rm{Q}} = 0.11\ \rm{K}$ and $T_{\rm{c}} = 0.05\ \rm{K}$, respectively \cite{Onimaru2010, Ishii.2}. The coexistence of superconductivity and quadrupolar order was also observed in $\rm{PrRh_2Zn_{20}}$, $\rm{PrTi_2Al_{20}}$, and $\rm{PrV_2Al_{20}}$ \cite{OnimaruPRB, Sakai.2, Tsujimoto}. The relationship between the superconductivity and the quadrupolar ordering has been intensively investigated \cite{Sakai.1, Umeo}. Non-Fermi-liquid (NFL) behavior was observed in specific-heat and electrical-resistivity measurements of $\rm{PrIr_2Zn_{20}}$ and its Y-diluted systems \cite{Onimaru2016, Yamane.1}. Since the NFL behavior may originate from the QKE, the material has attracted considerable attention.

 To search for the single-site effect of the QKE, some of the authors studied the Y-diluted compound  $\rm{Y_{0.966}Pr_{0.034}Ir_{2}Zn_{20}}$ (Pr-3.4\% system) by means of ultrasound, where they observed a logarithmic temperature dependence of the elastic constant $\left(C_{11}-C_{12}\right)/2$ below 0.3 K. This logarithmic behavior is a rare piece of experimental evidence to capture the single-site QKE in terms of quadrupolar susceptibility \cite{Yanagisawa.1}. In contrast, a recent theory of the QKE for a diluted system, assuming a virtual lattice after the random average over the positions of the quadrupolar ion  ($\rm{Pr^{3+}}$ ion), predicts that the quadrupolar susceptibility should be proportional to $\sqrt{T}$ at low temperatures even when the concentration of $\rm{Pr^{3+}}$ ion is only a few percent \cite{Tsuruta.1,Tsuruta.2}. In order to understand better the elastic response due to the QKE at low temperature, a systematic study of the non-Kramers systems $\rm{Y_{1-\it{x}}Pr_{\it{x}}Ir_{2}Zn_{20}}$ for a wide range of Pr concentrations is required. The present study, therefore, deals with ultrasonic measurements on $\rm{Y_{0.63}Pr_{0.37}Ir_{2}Zn_{20}}$ (Pr-37\% system) to investigate possible systematic changes of the elastic behavior. In the present paper, we report that the elastic constant $\left(C_{11}-C_{12}\right)/2$ of the Pr-37\% system shows a $\sqrt{T}$ temperature dependence below 0.15 K. 
 
 Moreover, we find a low-energy ultrasonic dispersion in the temperature range between 0.15 and 1 K. This suggests the existence of a low-energy phonon excitation, which has an energy scale close to the characteristic temperature $T^*\sim 1\ {\rm K}$ of the QKE. Thereby, entangled quantum states may be formed not only by the QKE but also by the electron-phonon coupling.

\begin{figure}
  \begin{center}
\includegraphics[clip,width=8cm]{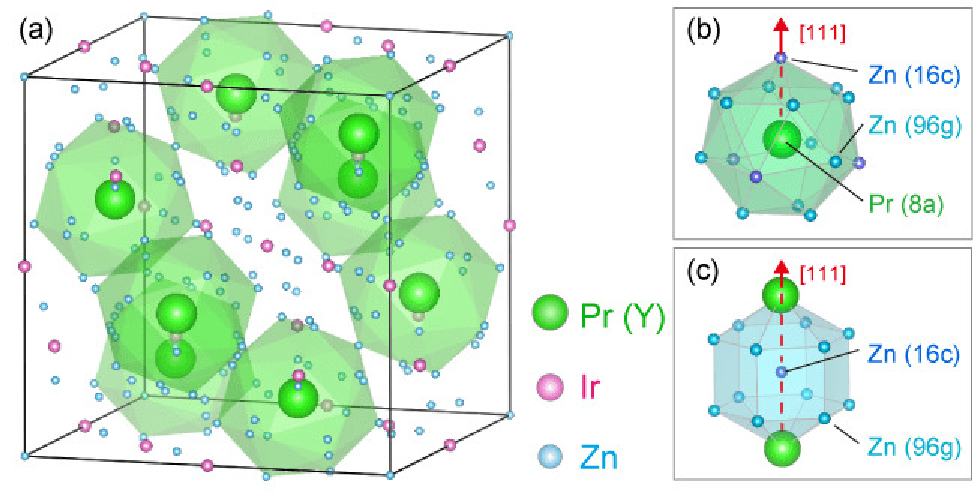}
\caption{  Crystal structure of $\rm{Y_{1-\it{x}}Pr_{\it{x}}Ir_{2}Zn_{20}}$. (a) A unit cell (the number of formula units per unit cell is 8), (b) a Pr atom encapsulated in the cage of Zn atoms (the site 16c and 96g), and (c) a Zn atom at the site 16c encapsulated in the cage of $\rm{Pr_2Zn_{12}}$, where the direction connecting the Pr atom (on site 8a) and the Zn atom (on site 16c) is [111]. These illustrations are drawn by using VESTA \cite{VESTA}
.\vspace{-7mm}}
\label{Fig_CS}
\end{center}
\end{figure}

\section{Experiment}

 The single crystals of $\rm{Y_{1-\it{x}}Pr_{\it{x}}Ir_{2}Zn_{20}}$ used in the present study were grown by the Zn-self-flux method with pre-arc-melting alloys of Y, Pr, and Ir as described in previous papers \cite{Yamane.1,Yamane.2}. The Pr composition $x = 0.37$ of the present sample was confirmed by the use of an electron probe micro analysis (EPMA) and CEF analysis of low-temperature magnetization measurements. The sample dimensions of the rectangular parallelepiped shape are $2.498\times1.813\times1.980\ \rm{mm^3}$ for $\left[110\right]-[1\bar{1}0]-\left[001\right]$. Ultrasound is generated and detected by a pair of $\rm{LiNbO_3}$ transducers with a thickness of 100 \textmu m, which were bonded on the well-polished sample surfaces with room-temperature-vulcanizing silicone. The quadrupolar responses can be observed as sound-velocity change by use of a phase-comparative method. The elastic constant $C_{ij}$ is converted from the sound velocity $v_{ij}$ by using the formula $C_{ij}=\rho{v_{ij}}^2$. Here, $ \rho=8.314\ \rm{g/{cm}^3}$ is the density of the present sample estimated from the lattice constant $a = 14.2234\ \AA$ for the Pr-37\% system. In the present measurements for $T >$ 2.3 K, the temperature was controlled by using a physical property measurements system (Quantum Design PPMS). The data for $T <$ 2.3 K were obtained by using a ${\rm^{3}He}$ refrigerator and a ${\rm^{3}He}$-${\rm^{4}He}$ dilution refrigerator.

\vspace{2mm}
\section{Results and Discussion}
\vspace{2mm}
\subsection{1. Ultrasonic Dispersion at around 40 K (UD1)}
\vspace{-3mm}
Figures 2(a) and (b)\ show the temperature dependence of the elastic constants $C_v=\left(C_{11}-C_{12}\right)/2$ and $C_{44}$, respectively, of the Pr-37\% system. The data are normalized and presented as relative change using absolute values at $2.4\ {\rm K}: C_{v}=4.64\times{10}^{10}\ \rm{J/m^3}$ and $C_{44}=4.45\times{10}^{10}\ \rm{J/m^3}$. $C_v$ shows a remarkable frequency-dependent upturn at around 40 K in the frequency range between 18 and 110 MHz and Curie-like softening below $5\ \rm{K}$. As shown in the lower part of Fig. 2(a), the background-subtracted ultrasonic attenuation coefficient exhibits a local maximum at which the elastic constant $C_v$ shows the upturn. The maximum shifts to higher temperatures with increasing ultrasonic frequencies. Here, a frequency-dependent background $\Delta\alpha_{0}\left(\omega, T\right)$ exhibiting monotonic increase with temperature is subtracted. On the other hand, $C_{44}$ displayed in Fig. 2(b)\ increases monotonically on cooling at least down to 2.3 K with no obvious frequency dependence.
 These mode-selective frequency dependences are reminiscent of the ultrasonic dispersion (UD) found in cage compounds such as $\rm{\it{R}_{\rm{3}}\rm{Pd_{20}Ge_{6}}}$ ($R$ = rare earth) \cite{Nemoto} and the filled-skutterudites $\rm{\it{R}\it{T}_{\rm{4}}\it{X}_{\rm{12}}}$ ($T$= Os, Ru, and Fe; $X$= P, Sb, and As) \cite{Goto,Yanagisawa.2,Ishii.0}. The origin of these UDs is considered as a ``rattling'' motion due to the cage-like crystal structure \cite{Goto}. Here, the rattling means a thermally activated oscillation of the guest atom in an oversized atomic cage \cite{Braun, Keppen}.\ Pr atoms at the site 8a and Zn atoms at the site 16c in the present $\rm{Y_{1-\it{x}}Pr_{\it{x}}Ir_{2}Zn_{20}}$ system are surrounded by highly symmetric cages as shown in Figs. 1(b) and 1(c). \  
 For the analysis of the UD in $C\left(\omega, T\right)$ and $\alpha\left(\omega, T\right)$, we adopt a Debye-type dispersion \cite{Nemoto} as
\begin{eqnarray}
C\left(\omega, T\right)\ &=&\ C\left(\infty, T\right)\ -\ \frac{C\left(\infty, T\right)-C\left(0, T\right)}{1+\omega^2\tau^2},\\
\alpha\left(\omega, T\right)\ &=&\ \frac{C\left(\infty, T\right)-C\left(0, T\right)}{2\rho{v_\infty}^3}\frac{\omega^2\tau}{1+\omega^2\tau^2}.
\end{eqnarray}

\begin{figure}
  \begin{center}
\includegraphics[clip,width=8.3cm]{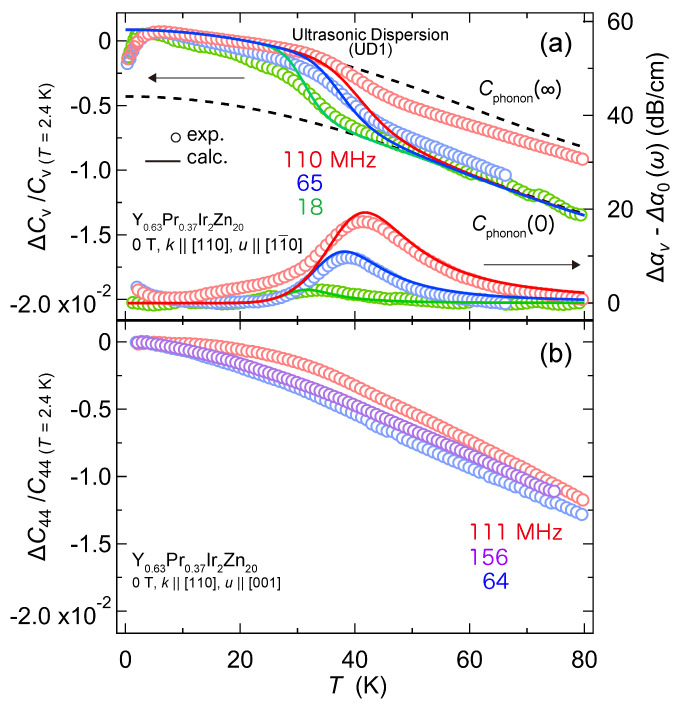}
\caption{ 
Relative change of the elastic constants (a) $C_v = \left(C_{11}-C_{12}\right)/2$ and (b) $C_{44}$ of the Pr-37\% system as a function of temperature. The ultrasonic attenuation coefficients for the $C_v = \left(C_{11}-C_{12}\right)/2$ mode are shown in the same panel (a). Three different measurement frequencies are shown in different colors. Black-dashed lines in panel (a) indicate the phonon background of the high- and low-frequency limit, $C_{\rm{phonon}}(\infty, T)$ and $C_{\rm{phonon}}(0, T)$, respectively. Solid lines are the calculation results analyzed based on a Debye-type dispersion to reproduce UD1 (see text).\vspace{-7mm}}
\label{Fig1}
\end{center}
\end{figure}
\begin{table*}[t]
 \caption{Ultrasonic dispersions observed in the Pr-based cubic systems. Resonant temperature means the temperature which satisfies the resonant equation ($\omega\tau\left(T\right) =1$) at around $\omega\ =\ 100$ MHz. The parameters for the UD of $\rm{PrIr_{2}Zn_{20}}$ (Pr-100\% system) are estimated by us from the data described in the original paper \cite{Ishii.2}. }
 \label{table:UD}
 \centering
  \begin{tabular}{cccccc}
   \hline
   compounds &resonant temperature (K) & activation energy $E$\ (K) & characteristic relaxation time $\tau_0$\ (s) & Refs.\\
   \hline
   $\rm{PrIr_{2}Zn_{20}}$ (Pr-100\%)  & 2 & $\rm{\sim\ 10}$ &$\sim\ 2.0\rm{\times\ 10^{-10}}$&[13]\\
   $\rm{Y_{0.63}Pr_{0.37}Ir_{2}Zn_{20}}$ (Pr-37\%) & 40 (UD1) & 250 &2.4$\rm{\times\ 10^{-11}}$& -\\
     &  0.5 (UD2) & 0.55 &3.1$\rm{\times\ 10^{-9}}$& -\\
   $\rm{Y_{0.966}Pr_{0.034}Ir_{2}Zn_{20}}$ (Pr-3.4\%) & - & -&-&[21]\\
   \hline
   $\rm{PrRh_{2}Zn_{20}}$  & 50 (UDH) & 440 & 2.0$\rm{\times\ 10^{-13}}$ &[32]\\
        &  2 (UDL)& 2.50 &8.0$\rm{\times\ 10^{-10}}$&[32]\\
   \hline
   \hline
  $\rm{PrOs_{4}Sb_{12}}$ & 35 & 168 &$8.8\rm{\times\ 10^{-11}}$&[27]\\
    \hline
   \hline
  $\rm{PrMg_{3}}$ & 0.2 & 0.50 &$8.5\rm{\times\ 10^{-10}}$&[33]\\
      \hline
  \end{tabular}
\end{table*}

 Here, $\omega$ is the angular frequency of the ultrasonic wave and $\tau$ is the relaxation time of the system. $C\left(\infty, T\right) $ and $C\left(0, T\right)$ denote the high- and low-frequency limits for the frequency-dependent elastic constant $C_v\left(\omega, T\right)$, respectively. Since the contribution of the phonons is obviously large compared to that of the CEF excitation at around 40 K,  we estimated that $C_{\rm{phonon}}\left(0, T\right)=4.600\times{10}^{10}-9.28T^2+4.176T^4\ \rm{J/m^3}$ from the experimental data and $C_{\rm{phonon}}\left(\infty, T\right)$ is vertically shifted by $0.025\times{10}^{10}\ \rm{J/m^3}$ as shown with the broken lines in Fig. 2(a). The colored solid lines in Fig. 2(a)\ are the results of a calculation assuming an Arrhenius-type relaxation represented by $\tau\left(T\right)=\tau_0\rm{exp}\left(\it{E/k_BT}\right)$, which roughly reproduces the experimental results. We obtained a characteristic relaxation time $\tau_0=2.4\times{10}^{-11}\ \rm{s}$ and an activation energy $E=250\ \rm{K}$ from the Arrhenius plot as shown in Fig. 3(b). 
 
 In addition, we investigated the magnetic-field dependence of UD1. Figure 3(a) exhibits results for $C_v$ obtained at selected frequencies (19, 65, and 109 MHz) and magnetic field of $B$ = 8 T. As shown in Fig. 3(b), the characteristic relaxation time and the activation energy are independent of the magnetic field, which is ascribed to the lattice vibrations.

\begin{figure}
  \begin{center}
    \includegraphics[clip,width=8.6cm]{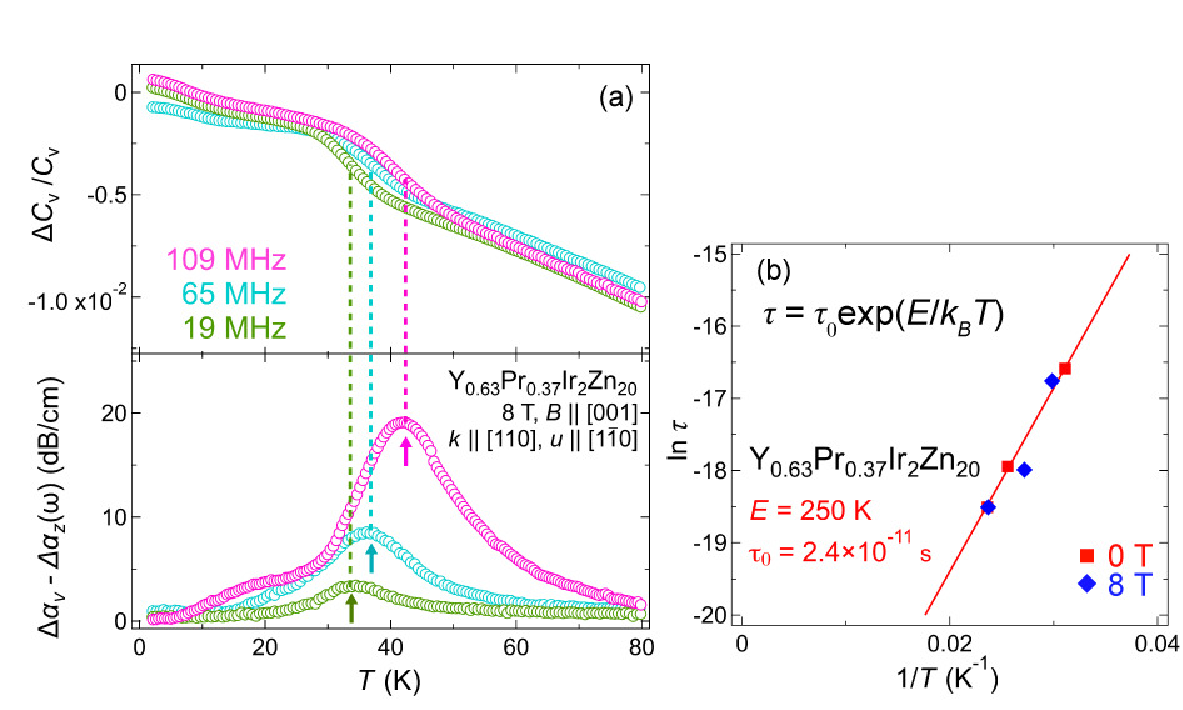}
    \caption{
(a) Relative change of the elastic constant $C_v \left(= \left(C_{11}-C_{12}\right)/2\right)$ and the ultrasonic attenuation coefficient of the Pr-37\% system at 8 T $\left(B\ ||\ [001]\right)$ as a function of temperature. The results at three different measurement frequencies are shown in different colors. Each arrow shows the resonant point ($\omega\tau=1$). (b) Arrhenius plot of the logarithm of the relaxation time $\rm{ln}\tau$ vs the reciprocal temperature $1/T$ for UD1 at $B$ = 0 (red) and 8 T (blue). The solid line shows the fitted straight line to the data in $B$ = 0.}
  \end{center}
\end{figure}
 
 As shown in the previous report by Ishii $et\ al$. \cite{Ishii.2}, a UD at around 2 K was observed in the Pr-100\% system. The activation energy of this UD was estimated to be about 10 K. Assuming that the origin of this UD is same as for UD1 of the Pr-37\% system,  the energy scale of the UD decreases as the Pr concentration increases. Previous powder x-ray diffraction measurements pointed out that the lattice constant of $\rm{Y_{1-\it{x}}Pr_{\it{x}}Ir_{2}Zn_{20}}$ increases as the Pr concentration increases \cite{Yamane.2}. Thus, this Pr concentration dependence can be interpreted as a size increase of the atomic cage together with an energy decrease of the vibration as the Pr concentration increases.  Since no UD was observed for the Pr-3.4\% system below 150 K \cite{Yanagisawa.1}, it may be possible that such a UD exists at higher temperatures for the Pr-3.4\% system. However, the UD in the Pr-100\% system reported by Ishii $et\ al$. shows some differences from that of the Pr-37\% system, such as that UD is observed for both $C_v$ and $C_{44}$ modes.

On the other hand, a mode-selective UD was observed previously for $\rm{PrRh_2Zn_{20}}$, having the same crystal structure as $\rm{PrIr_2Zn_{20}}$ \cite{Ishii.1}. This UD is named as ``UDH" in the original paper \cite{Ishii.1}. ($\rm{PrRh_2Zn_{20}}$ has the other UD named as ``UDL", which has a similar energy scale to UD2 described in the section 3.3 later.) The activation energy of UDH is $\ \it{E}=\rm{440\ K}$, which is close to the present values of UD1 ($\it{E}=\rm{250\ K}$). Recently, low-energy optical-phonon excitations at $\sim\rm{7\ meV}$ were observed by inelastic x-ray scattering (IXS) in $\rm{Pr}\it{T}_2\rm{Zn}_{20} \left(\it{T} = \rm{Rh, Ir}\right)$. It was concluded that such low-energy phonon modes originate from the oscillation of Zn atoms (site 16c) encapsulated in the $\rm{Pr_2Zn_{12}}$ cage (shown in Fig. 1(c)) \cite{Wakiya.1, Wakiya.2}. 
First-principal calculations for LaRu$_2$Zn$_{20}$ (same structure as PrIr$_2$Zn$_{20}$) suggests that the vibrations of the $\rm{\Gamma_3}\ ({\it E}_{{\rm g}})$ and $\rm{\Gamma_5}\ ({\it T}_{{\rm 2g}})$ modes in the (111) plane have very low energies. As their frequencies are so low and sensitive to the parameters of the calculation, thus, the symmetry of vibrations has not been revealed yet \cite{Hasegawa}.
The present ultrasonic results suggest that this vibration corresponds to the $\rm{\Gamma_3}$ mode. Further experiments are needed to clarify the origin of the ultrasonic dispersion.


\begin{figure}
  \begin{center}
\includegraphics[clip,width=7.8cm]{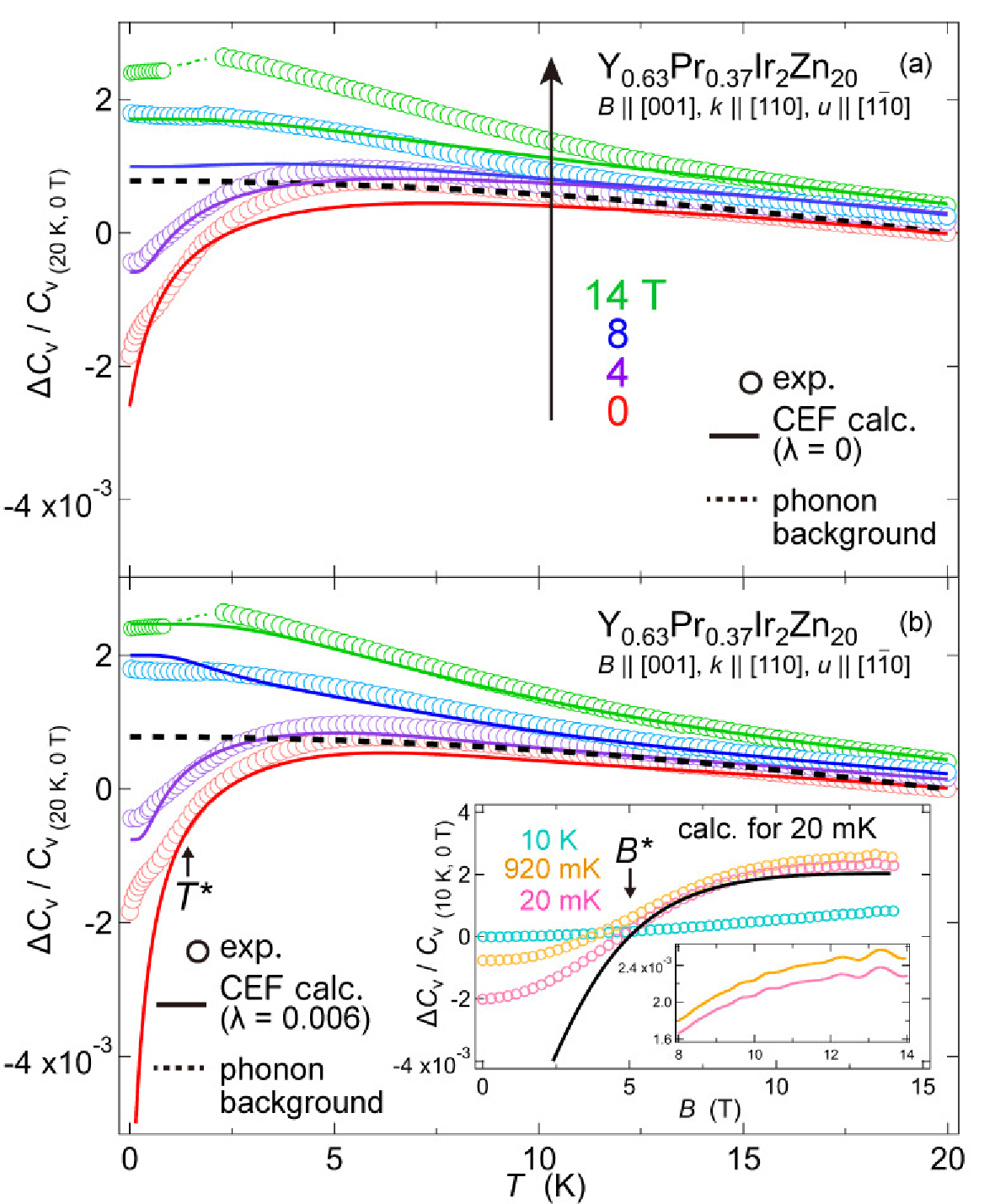}
\caption{Relative change of the elastic constant $C_v$ of the Pr-37\% system as a function of temperature at various magnetic fields $B\ ||\ \rm{[001]}$. Data above and below 0.5 K were measured with frequency of 110 and 201 MHz, respectively. Solid lines in (a) and (b) show the calculation results, respectively, without and with the contribution of the electric hexadecapole $H_v$ ($\lambda\ =\ 0.006$). The black broken line indicates the background elastic constant (high-frequency limit). The inset in (b) shows the relative change of $C_v$ as a function of magnetic field at $T=0.020, 0.920,$ and 10 K. The black solid curve in the inset is the  CEF calculation for 20 mK. $B^*$ is the characteristic magnetic field (see text). In addition, magneto-acoustic quantum oscillations were observed above 9 T at 20 and 920 mK (as shown in the enlarged view).\vspace{-7mm}}
\label{Fig2}
\end{center}
\end{figure}

\subsection{2. Crystalline-Electric-Field Analysis with Electric-Hexadecapole Contribution}
\vspace{-3mm}

 The main panel of Fig. 4\ shows the temperature dependences of $C_v$ at various magnetic fields $B\ ||\ \rm{[001]}$ below 20 K. These temperature dependences of $C_v$ can be analyzed by localized CEF models with the following Hamiltonian \cite{LLW,Hutchings}, $\mathcal{H}=\mathcal{H}_{\rm{CEF}}+\mathcal{H}_{\rm{Zeeman}}+\mathcal{H}_{\rm{MS}}+\mathcal{H}_{\rm{MM}}$ (see details in the Appendix).

In principle, it is necessary to consider the hyperfine interaction between nuclear spins and localized quadrupoles, since the nuclear spin of the $\rm{^{141}Pr}$ nucleus (natural abundance is 100\%) is non-zero ($I$ = 5/2).
In our previous study, we took into account the effect of hyperfine interactions, which could explain the appearance of a local minimum of the elastic constant at low magnetic fields \cite{Yanagisawa.1}.
The present results do not show such a local minimum. In fact, our calculations without hyperfine interaction already reproduce the experimental results well (as described in detail later), therefore, we conclude that the contribution of the hyperfine interaction is sufficiently small to be negligible at least in the present system.

 We use the CEF parameters for $\rm{PrIr_2Zn_{20}}$: $B_4^0=-0.0109\ \rm{K}$ and $B_6^0=-0.4477\ \rm{mK}$ in the present CEF Hamiltonian $\mathcal{H}_{\rm{CEF}}$ for the Pr-37\% sample, since they as well can be used for the analysis for the ultrasonic data of the Pr-3.4\% sample \cite{Iwasa,Yanagisawa.1}.  
 
 The CEF level scheme based on these parameters is as follows; $\rm{\Gamma_3}\ (0\ K)$-$\rm{\Gamma_4}\ (27\ K)$-$\rm{\Gamma_1}\ (66\ K)$-$\rm{\Gamma_5}\ (67\ K)$. 
 The third term of the Hamiltonian describes the interaction between the elastic strain and an effective multipole, and the fourth term describes the intersite multipole-mulipole interaction. 
 Under the assumption with an effective multipole ${\widetilde{O}}_v=O_v+{\lambda H}_v$ with $\Gamma_3$ symmetry, we compare the results with and without the contribution of $H_v$ in Figs. 4(a) and (b). 
 Here, $O_v=J_x^2-J_y^2$ is the $\Gamma_3$-type electric quadrupole and $H_v=1/4[\{7J_z^2-J\left(J+1\right)-5\}\left(J_+^2+J_-^2\right)+\left(J_+^2-\ J_-^2\right)\{7J_z^2-J\left(J+1\right)-5\}]$ is the $\Gamma_3$-type electric hexadecapole. $\lambda$ means the contribution ratio of the hexadecapole to the quadrupole. The calculation with $\lambda=0.006$ (Fig. 4(b)) reproduces the experimental data much better than the calculation without including the electric hexadecapolar contribution (Fig. 4(a)), in particular, in the range of $B > 8$ T.  Thereby, it is necessary to consider not only the contribution of the electric quadrupole $O_v$ but also that of the electric hexadecapole $H_v$ to reproduce the data. 

Here, Such non-negligible contribution of the electric hexadecapole ($\lambda\neq\ 0$) has already been demonstrated by ultrasonic measurements of Pr systems with a $\Gamma_3$-$\Gamma_4$ state \cite{Araki.1}. 
The elastic constant $C_{\mathrm{\Gamma}_3}^{4f}(T,\ B)$ can be expressed by the following formula \cite{Luthi}:
\begin{eqnarray}
C_{\mathrm{\Gamma}_3}^{4f}(T,\ B)=C_{\mathrm{\Gamma}_3}^0(T)-\frac{Ng_{\mathrm{\Gamma}_3}^2\chi_{\mathrm{\Gamma}_3}(T,\ B)}{1-{g^{\prime}}_{\mathrm{\Gamma}_3}\chi_{\mathrm{\Gamma}_3}(T,\ B)}.
\end{eqnarray}
Here, 
$N=8x/a(x)^3=1.03\times{10}^{27}\ \rm{m^{-3}}$ is the number of Pr ions per unit volume in the Pr-37\% sample, estimated from the lattice constant $a(x)$ at room temperature,

and $C_{\mathrm{\Gamma}_3}^0$ is the background elastic constant which originates from the phonon contribution ($C_{\rm{phonon}}\left(\infty, T\right)$, black broken line in Fig. 4). $g_{\mathrm{\Gamma}_3}$ and $g_{\mathrm{\Gamma}_3}^{\prime}$ are the coupling constants of the multipole-strain interaction and the multipole-multipole interaction, respectively. Then, $\chi_{\Gamma_3}$ is the susceptibility of the effective multipole ${\widetilde{O}}_v$. 

  The coupling constants, $g_{\mathrm{\Gamma}_3}=40\ \rm{K}$ and $g^{\prime}_{\mathrm{\Gamma}_3}=-0.05\ \rm{K}$ are obtained from the fit to the data at various magnetic fields above 1 K, since a QKE and a change in the ground state due to a low-energy phonon excitations affect the data at lower temperatures (as described in detail below). The present parameter set for the calculation reproduces the magnetic-field dependence above $5\ \rm{T}$, as shown in the inset of Fig. 4(b). 
Here, we define the characteristic magnetic field $B^*$ below which the localized 4$f$-electron model is no longer valid. $B^*$ can be estimated as $5\ \rm{T}$ at the lowest temperature of 20 mK. It should be noted that magneto-acoustic quantum oscillations were observed above 9 T at 20 and 920 mK. This phenomenon proves the high quality of the present single crystal.

  In addition, we reanalyzed previously reported data for the Pr-3.4\% \cite{Yanagisawa.1} system and the Pr-100\% \cite{Ishii.2} system in zero field by considering hexadecapolar contributions using the same parameter $\lambda=0.006$ (Fig. 5). In case of the Pr-3.4\% system, $g'$ is equal to zero, indicating that the contribution of the electric hexadecapole is negligible. On the other hand, the electric hexadecapolar contribution cannot be ignored in the Pr-100\% system. The analysis as well $\lambda\ =\ 0.006$ reproduces the data well for all Pr concentrations.  Figure 6(a)\ shows $g_{\mathrm{\Gamma}_3}$ obtained with $\lambda = 0.006$ (calc. 1) and $\lambda = 0$ (calc. 2) for each Pr concentration. Figure 6(b)\ shows $g^{\prime}_{\mathrm{\Gamma}_3}$ correspondingly. The hexadecapolar contribution suppresses the Curie-like softening. Thus, the values of $g_{\mathrm{\Gamma}_3}$ are larger than those considering only the contribution of the electric quadrupole. The value of $g_{\mathrm{\Gamma}_3}$ in the present Pr-37\% system is $\sim1.5$ times larger than for $\lambda = 0$. In  any case, the absolute values of both $g_{\mathrm{\Gamma}_3}$ and $g^{\prime}_{\mathrm{\Gamma}_3}$ increase linearly with Pr concentration.

 \begin{figure}[t]
  \begin{center}
    \includegraphics[clip,width=8.2cm]{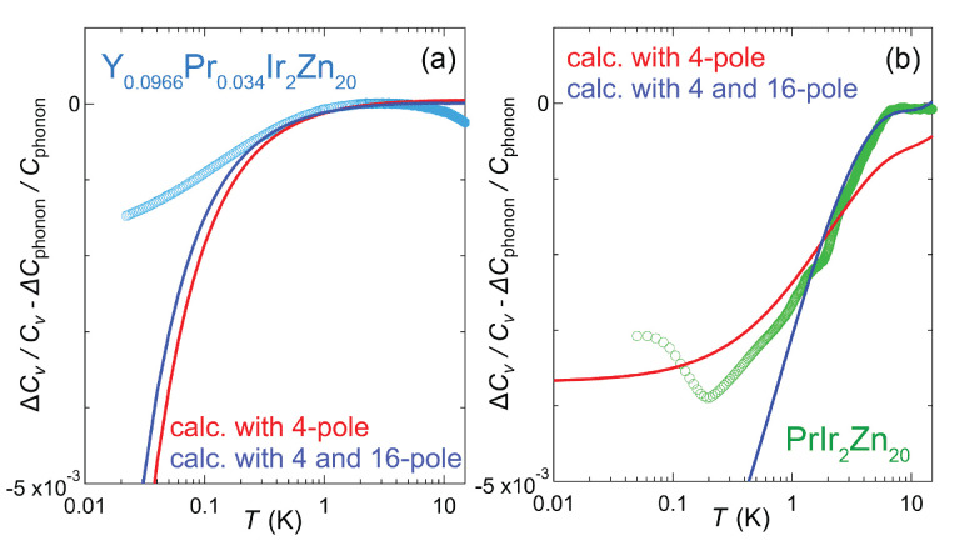}
    \caption{Relative change of the elastic constants $C_v$ with phonon background subtracted of the Pr-3.4\% system \cite{Ishii.2} (left) and the Pr-100\% system \cite{Yanagisawa.1} (right) as a function of temperature. The red and blue lines show the calculations without and with the contribution of the electric hexadecapole $H_v$ ($\lambda=0.006$). }
  \end{center}
\end{figure}

\subsection{3. Ultrasonic Dispersion at around 0.5 K (UD2)}
\vspace{-3mm}

 Figure 7\ shows the temperature dependence of the elastic constant $C_v$  at the frequencies of 65, 110, and 201 MHz below 10 K \cite{enviroment}. The purple dashed line at higher temperatures is obtained using the CEF calculation as described above. $C_v$ at 110 MHz deviates from the calculation below $T^*\sim1\ \rm{K}$. Here, we defined $T^*$ as the characteristic temperature below which the localized CEF model seems to be no longer applicable.
 
 In the present Pr-37\% system, $C_v$ shows frequency dependences at around 0.5 K, while $C_{44}$ does not show a frequency dependence also at this low-temperature region (not shown). It is different from that $C_v$ in the Pr-3.4\% system shows no frequency dependence down to 0.04 K in the frequency range from 63 to 237 MHz \cite{Yanagisawa.1}. 
 
\begin{figure}
\begin{minipage}{\hsize}
 \begin{center}
\vspace{3mm}
\includegraphics[clip,width=7.7cm]{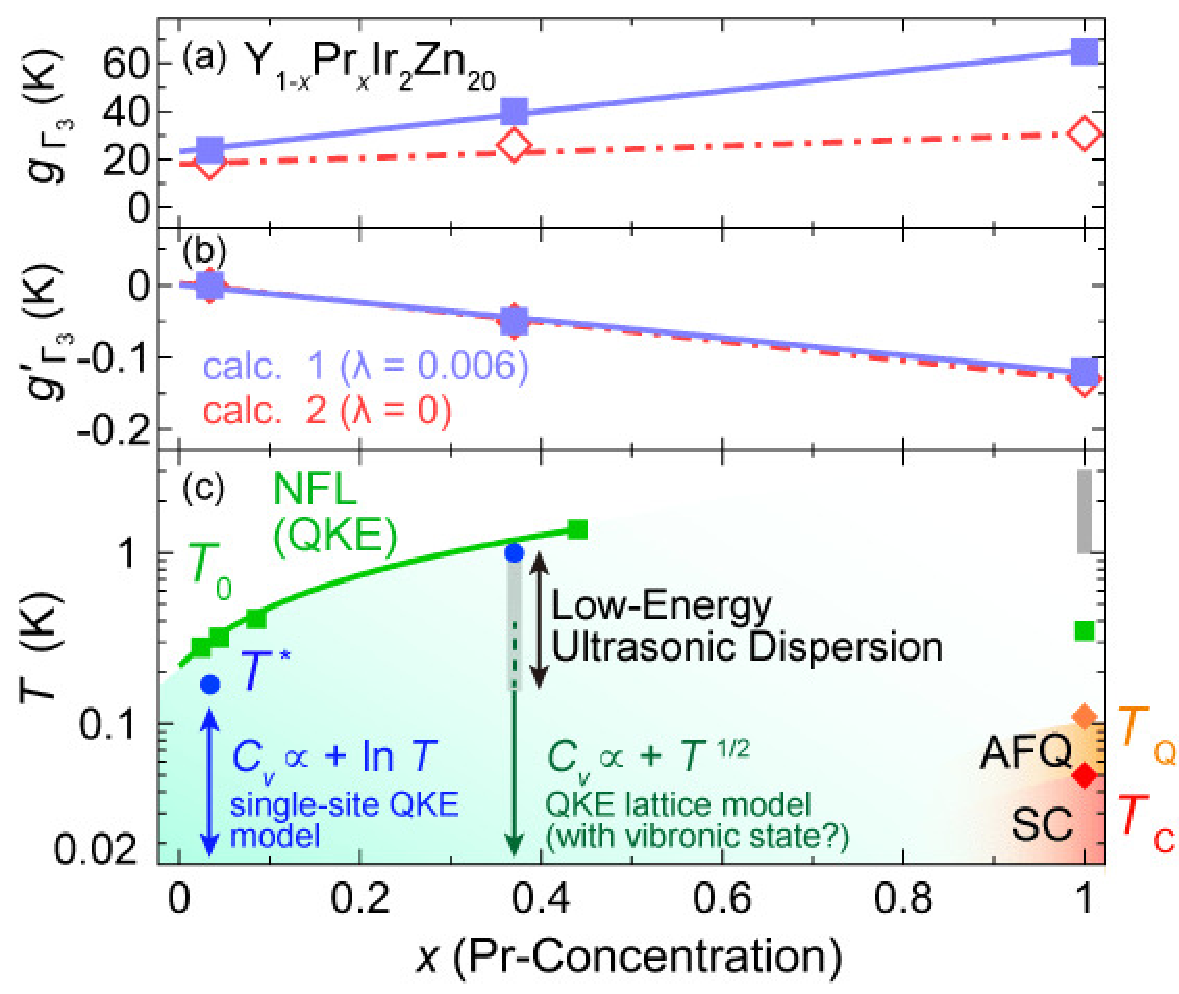}
\caption{The coupling constants of (a) the multipole-strain interaction $g_{\mathrm{\Gamma}_3}$ and (b) the multipole-multipole interaction $g_{\mathrm{\Gamma}_3}^{\prime}$ as a function of Pr concentration $x$ for $\rm{Y_{1-\it{x}}Pr_{\it{x}}Ir_{2}Zn_{20}}$ \cite{Yanagisawa.1, Ishii.2}. Blue closed markers (calc. 1) and red open diamonds (calc. 2) are parameters obtained by fits with ($\lambda=0.006$) and without ($\lambda=0$) electric hexadecapolar contribution, respectively. (c) $T$-$x$ phase diagram of $\rm{Y_{1-\it{x}}Pr_{\it{x}}Ir_{2}Zn_{20}}$. $T_0$ is the characteristic temperature of the QKE determined from specific-heat experiments \cite{Yamane.1}. $T^*$ is the temperature at which the behavior of the ultrasound mode $C_v$ deviates from the CEF model \cite{Yanagisawa.1}.}
\label{Fig4}
\end{center}
\end{minipage}
\begin{minipage}{\hsize}
  \begin{center}
\vspace{3mm}
\includegraphics[clip,width=8cm]{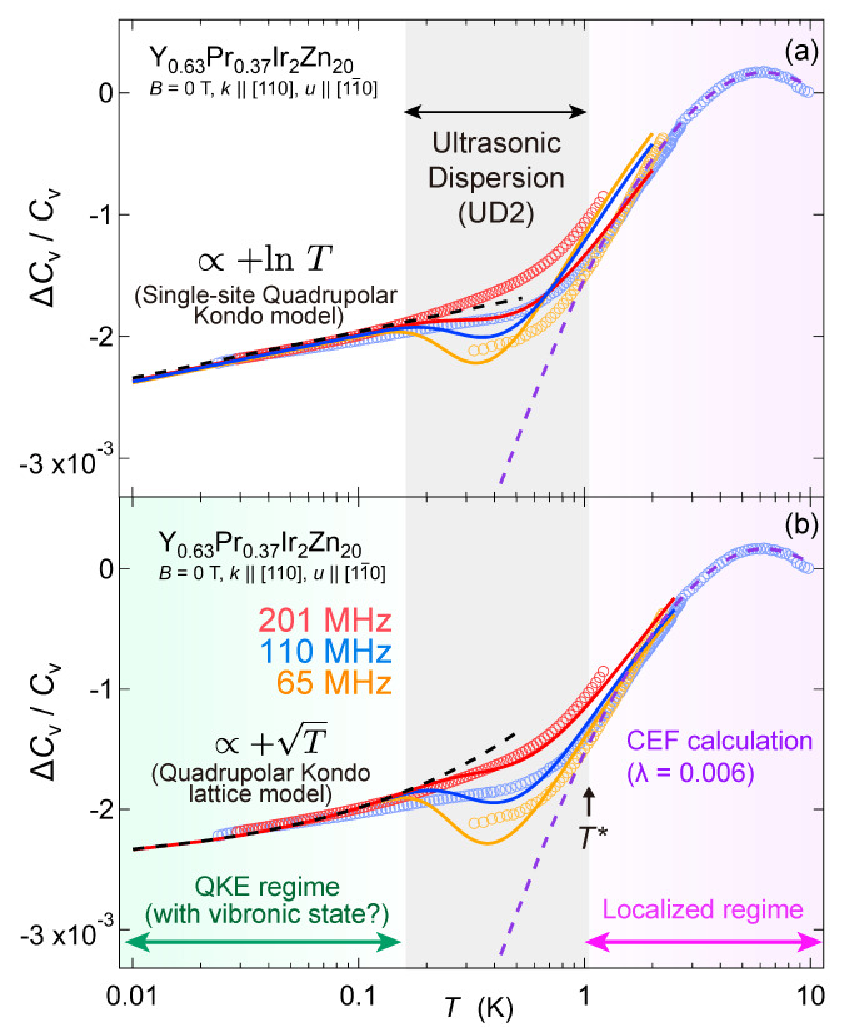}
\caption{Relative change of the elastic constants $C_v$ of the Pr-37\% system as a function of temperature below 10 K. The purple broken lines show the CEF calculation using $\lambda=0.006$. The black broken lines show the elastic constants estimated for the high-frequency limit, $C_v(\infty)$ (the line in (a) is proportional to ln $T$, though that in (b) follows $\sqrt{T}$). Solid lines are the calculations assuming a Debye-type dispersion to reproduce UD2.}
\label{Fig3}
\end{center}
\end{minipage}
\end{figure}

Here, for UD2, we apply the same Arrhenius-type analysis as for UD1 at high temperatures. The solid lines in Figs.7(a) and 7(b) show the results based again on a Debye-type dispersion (Eq. (1)). Here, we assume that the high-frequency limit $C\left(\infty, T\right)$ has (a) a logarithmic temperature dependence based on the single-site quadrupolar Kondo model and (b) a square-root temperature dependence based on the quadrupolar Kondo (virtual) lattice model \cite{Tsuruta.1}. For both calculations, we assume that the low-frequency limit $C\left(0, T\right)$ is calculated using the localized CEF model. The latter analysis reproduces the experimental data better, in particular, for the data at high frequency. This suggests that the quadrupolar Kondo (virtual) lattice model should be applied in the Pr-37\% system, where the interaction between localized quadrupoles is not negligible, unlike the Pr-3.4\% system.

We obtain the relatively slow characteristic relaxation time $\tau_0\ =\ 3.1\times{10}^{-9}\ \rm{s}$ and the low activation energy $E\ \approx\ 550\ \rm{mK}$. This energy scale is much smaller than that of UD1, indicating that UD1 and UD2 have different origins. 
 The energy scale of UD2 is as well much smaller than that of UD observed in the Pr-100\% system ($E\sim10\ {\rm K}$). On the other hand, a similar low-energy UD was observed in the non-Kramers compound $\rm{PrMg_3}$ ($\tau_0=0.85\times{10}^{-9}\ \rm{s},\ {\it E}=500\ \rm{mK}$) \cite{Araki.2}.

 $\rm{PrMg_3}$ shows no long-range quadrupolar order at least down to 20 mK \cite{Tanida}. Previously, Araki $et\ al.$ pointed out possible formation of a vibronic state, which is a quantum state caused by the coupling between the non-Kramers doublet and the phonons of the surrounding lattice near absolute zero \cite{Araki.2}. It has been theoretically predicted that a Kondo-like singlet can be formed by the coupling of the $\Gamma_3$ doublet to dynamical Jahn-Teller phonons \cite{Hotta.1,Hotta.2}.\ Based on these experimental and theoretical proposals and the similarities to the present case, it is possible that a vibronic state is formed in the NFL region of the Pr-37\% system. From the perspective shown in Fig. 1(b), a Pr atom is surrounded by the high-symmetry cage of sixteen Zn atoms, which suggests the existence of an isolated low-energy phonon at the Pr site (interpreted as Einstein phonon).\ Due to this structural situation, it is relatively easy to form a ``vibronic state" coupling between the low-energy phonons and the localized electric quadrupoles to release the entropy of the CEF ground-state doublet $\Gamma_3$.

\subsection{4. Dilution Effect on the Quantum State below the Characteristic Temperature $T^*$}
\vspace{-3mm}

As we discussed, the elastic response of $\rm{Y_{1-\it{x}}Pr_{\it{x}}Ir_{2}Zn_{20}}$ at very low temperature changes drastically depending on the Pr-concentration $x$. Figure 6(c)\ summarizes the change in the quantum state as a function of $x$. Here, $T_0$ is the characteristic temperature of NFL behavior, which is obtained from the specific-heat measurements \cite{Yamane.1}. $T^*$, the characteristic temperature of the QKE observed by our ultrasonic measurements, lies close to $T_0$.
In the Pr-37\% system, it is possible to form a vibronic state, whereas a square-root temperature dependence of $C_v$ expected for the quadrupolar Kondo lattice model was observed below 0.15 K. The present results suggest that \ the Pr-Pr\ interactions become non-negligible with increasing $x$, leading to a crossover from a quantum ground state of single-site QKE to a quadrupolar Kondo lattice, which may occur for Pr concentrations between 3.4 \% and 37 \%.

In the present paper, we analyze the temperature dependence of $C_v$ at very low temperatures assuming that UD2 originating from electron-phonon coupling and the temperature dependence proportional to $\sqrt{T}$, caused by the quadrupolar Kondo (virtual) lattice effect are independent. Such analysis qualitatively explains the present experimental results. However, whether these quantum states are independent or entangled is still an open question. To address the issue, it is necessary to construct a new theory including a localized quadrupolar moment and conduction electrons as well as phonons to describe the quantum ground state of $\rm{Y_{1-\it{x}}Pr_{\it{x}}Ir_{2}Zn_{20}}$. Therefore, it would be highly desirable to investigate the Pr concentration dependence in more detail through a series of experiments on $\rm{Y_{1-\it{x}}Pr_{\it{x}}Ir_{2}Zn_{20}}$.


%
%

  Finally, we discuss the multiple effects of the elemental dilution. 
The Y-dilution changes the following three factors, \newcounter{num}\setcounter{num}{1}\roman{num}) the number density of the localized Pr moments, \setcounter{num}{2}\roman{num}) the lattice constant, and \setcounter{num}{3}\roman{num}) the atomic disorder. These three contributions can affect the magnitude of the Ruderman-Kittel-Kasuya-Yosida (RKKY)-type interaction among the localized quadrupoles. As the Pr concentration decreases, the lattice constant decreases \cite{Yamane.2}. However, as the AFQ order observed in the Pr-100\% system collapses in Y-containing systems, the effective RKKY-type interaction between the Pr$^{3+}$ ions becomes weakened with the Y dilution. Indeed, the absolute value of the interaction between multipoles $g^{\prime}_{\mathrm{\Gamma}_3}$ (Fig. 6(b)) decreases linearly with decreasing Pr concentration. 
 Then, the characteristic temperature of the QKE, $T^*$ (or $T_0$) increases as the Pr concentration increases up to 37\%. We assume that the change of $T^*$ is ascribed to the atomic disorder which loweres the site symmetry of $\rm{Pr^{3+}}$ and locally splits the ground-state doublet $\Gamma_3$. Since the atomic disorder is generally largest for $x=0.5$ (Y : Pr = 1 : 1), this assumption is consistent with the Pr-concentration dependence of the characteristic temperatures.  
It is necessary to investigate the Pr-concentration dependence of the energy scale of the low-energy UD and compare it with that of the QKE characteristic temperature, which would serve for understanding the relationship between the QKE and the vibronic ground state.
 
 \section{Conclusion}
In summary, the elastic constant $\left(C_{11}-C_{12}\right)/2$ of $\rm{Y_{0.63}Pr_{0.37}Ir_{2}Zn_{20}}$ (Pr-37\% system) shows a temperature dependence proportional to $\sqrt{T}$ below 0.15 K, which corresponds to the theoretical expectation for a quadrupolar Kondo ``lattice'' model. From a comparison with $\left(C_{11}-C_{12}\right)/2$ of the Pr-3.4\% system, which shows a logarithmic temperature dependence, we conclude that the crossover from the single-site model to the virtual-lattice model is caused by the increasing Pr concentration. At the Pr-37\% system, we find a low-energy phonon excitation with an energy scale which is close to that of the QKE characteristic temperature. Further experiments are necessary to check the possibility that this low-energy phonon excitation couples to the $\Gamma_3$ doublet and influences the simple QKE behavior at intermediate Pr concentrations. \\

\begin{acknowledgments}
The present research was supported by JSPS KAKENHI Grants Nos. JP15KK0169, JP18H04297, JP18H01182, JP17K05525, JP18KK0078, JP15KK0146, JP15H05882, JP15H05885, JP15H05886, JP15K21732, JP21KK0046, and JP22K03501. We acknowledge support from the DFG through the W\"{u}rzburg-Dresden Cluster of Excellence on Complexity and Topology in Quantum Matter - {\it ct.qmat} (EXC 2147, project-id 39085490) and from the HLD at HZDR, member of the European Magnetic Field Laboratory (EMFL). This study was partly supported by Hokkaido University, Global Facility Center (GFC), Advanced Physical Property Open Unit (APPOU), funded by MEXT under Support Program for Implementation of New Equipment Sharing System Grant No. JPMXS0420100318.
\end{acknowledgments}

\newpage
\begin{appendix}
\section{Appendix: Calculation of the Multipolar Susceptibility \\ for a Cubic CEF}

We show the calculation of the temperature and magnetic-field dependence of the elastic constant. Our calculation uses the theory based on the Brillouin-Wigner perturbation method \cite{Luthi}. As shown in the present paper, we assume that the Hamiltonian is
\footnotesize
\begin{equation}
\mathcal{H}\ = \mathcal{H}_{\rm{CEF}}+ \mathcal{H}_{\rm{Zeeman}}+ \mathcal{H}_{\rm{MS}}+ \mathcal{H}_{\rm{MM}}. \tag{A$\cdot$1}
\end{equation}
\normalsize
Here, the first term describes the CEF hamiltonian and the second term corresponds to Zeeman interaction.
The third term describes the interaction between the elastic strain and an effective multipole, and the fourth term describes the intersite multipole-mulipole interaction.
Since the site symmetry of $\rm{Pr^{3+}}$ is $T_d$ (cubic), the CEF Hamiltonian  $\mathcal{H}_{\rm{CEF}}$ can be expressed as follows using the two independent CEF parameters $B_4^0$ and $B_6^0$. Here, $O_m^n$ are the Stevens operators,
\footnotesize
\begin{equation}
\mathcal{H}_{\rm{CEF}}=B_4^0\left(O_4^0+5O_4^4\right)+B_6^0\left(O_6^0-21O_6^4\right). \tag{A$\cdot$2}
\end{equation}
\normalsize
We consider only the coupling Hamiltonian of the elastic strain to an effective multipole, $\mathcal{H}_{MS}\left(\Gamma\right)=-g_{\Gamma}\tilde{O}_{\Gamma}\varepsilon_{\Gamma}$ (third term of Eq. (A$\cdot$1)) as a perturbation Hamiltonian. $E_i\left(\varepsilon_\Gamma\right)$ is a perturbed CEF level as a function of strain $\varepsilon_\Gamma$ up to second-order perturbation, which can be written as
\footnotesize
\begin{align}
E_i\left(\varepsilon_\Gamma\right)\ &=\ E_i^0+\Braket{i|\mathcal{H}_{\rm{MS}}|i}+\sum_{j \neq i}\frac{\left|\Braket{j|\mathcal{H}_{\rm{MS}}|i}\right|^2}{E_j^{(0)}-E_i^{(0)}}\nonumber\\
&=\ E_i^0-g_\Gamma\Braket{i|\tilde{O}_\Gamma|i}\varepsilon_\Gamma+g_\Gamma^2\sum_{j \neq i}\frac{\left|\Braket{j|\tilde{O}_\Gamma|i}\right|^2}{E_j^{(0)}-E_i^{(0)}}\varepsilon_\Gamma^2. \tag{A$\cdot$3}
\end{align}
\normalsize
The Helmholtz free energy of the local $4f$-electron states in the CEF can be written as,
\footnotesize
\begin{align}
F\ &=\ U-Nk_BT\rm{ln}\sum_{\it{i}}\rm{exp}\left\{\it{-E_i\left(\varepsilon_{\Gamma}\right)/k_BT}\right\}\nonumber\\
&=\ \frac{1}{2}\sum_{\Gamma}C_\Gamma\varepsilon_\Gamma^2-Nk_BT\rm{ln}\sum_{\it{i}}\rm{exp}\left\{\it{-E_i\left(\varepsilon_\Gamma\right)/k_BT}\right\}, \tag{A$\cdot$4}
\end{align}
\normalsize
where $N$ is the number of ions in a unit volume, $i$ is the number index for $J$ multiplets and their degenerate states. $U=\frac{1}{2}\sum_{\Gamma}C_\Gamma\varepsilon_\Gamma^2$ is the internal energy for the strained system. Since the elastic constant with $\Gamma$-type symmetry, $C_\Gamma$, is defined as the second derivative of the free energy with respect to strain $\varepsilon_\Gamma$, $C_\Gamma$ is derived as follows from Eq. (A$\cdot$4):
\newline
\footnotesize
\begin{align}
\hspace{-5mm}C_\Gamma\left(T,B\right)&=\left(\frac{\partial^2F}{\partial\varepsilon_\Gamma^2}\right)_{\varepsilon_{\Gamma}\rightarrow0}\nonumber\\
&=C_\Gamma^0+N\left[\Braket{\frac{\partial^2E_i}{\partial\varepsilon_\Gamma^2}}-\frac{1}{k_BT}\left\{\Braket{\left(\frac{\partial E_i}{\partial\varepsilon_\Gamma}\right)^2}-\Braket{\frac{\partial E_i}{\partial\varepsilon_\Gamma}}^2\right\}\right]. \tag{A$\cdot$5}
\end{align}
\normalsize
Here, $C_\Gamma^0$ is the background of the elastic constant. The single-ion multipolar susceptibility $\chi_\Gamma$ is defined as 
\footnotesize
 \begin{equation}
 g_\Gamma^2\chi_\Gamma=-\Braket{\frac{\partial^2E_i}{\partial\varepsilon_\Gamma^2}}+\frac{1}{k_BT}\left[\Braket{\left(\frac{\partial E_i}{\partial\varepsilon_\Gamma}\right)^2}-\Braket{\frac{\partial E_i}{\partial\varepsilon_\Gamma}}^2\right]. \tag{A$\cdot$6}
 \end{equation}
 \normalsize
Thus, $C_\Gamma$ is described as follows from Eqs. (A$\cdot$5) and (A$\cdot$6):
\footnotesize
\begin{equation}
C_\Gamma\left(T,B\right)=C_\Gamma^0-Ng_\Gamma^2\chi_\Gamma\left(T,B\right). \tag{A$\cdot$7}
\end{equation}
\normalsize
In addition to the strain-multipole interaction $\mathcal{H}_{MS}$, the intersite multipole-multipole interaction can also be added by using a molecular-field approximation of the multipolar moment $\tilde{O}_\Gamma$ as
\footnotesize
\begin{equation}
\mathcal{H}_{MM}\left(\Gamma\right)=\sum_{\Gamma}g_{\Gamma}'\Braket{\tilde{O}_\Gamma} \tilde{O}_\Gamma. \tag{A$\cdot$8}
\end{equation}
\normalsize
\\
This corresponds to considering the effective strain $\varepsilon_\Gamma^{\rm{eff}}$ as
\footnotesize
\begin{align}
\mathcal{H}_{MS}+\mathcal{H}_{MM}\ &=\ \sum_{\Gamma}g_\Gamma\left(\varepsilon_\Gamma+\frac{g_{\Gamma}'}{g_{\Gamma}}\Braket{\tilde{O}_\Gamma}\right)\tilde{O}_\Gamma\nonumber\\
&=\ -\sum_{\Gamma}g_\Gamma\varepsilon_\Gamma^{\rm{eff}}\tilde{O}_\Gamma. \tag{A$\cdot$9}
\end{align}
\normalsize
The elastic constant $C_\Gamma$ is rewritten as
\footnotesize
\begin{equation}
C_\Gamma\left(T,B\right)=C_\Gamma^0-Ng_\Gamma^2\frac{\chi_\Gamma\left(T,B\right)}{1-g_\Gamma'\chi_\Gamma\left(T,B\right)}. \tag{A$\cdot$10}
\end{equation}
\normalsize
From Eqs. (A$\cdot$3) and (A$\cdot$6), the multipolar susceptibility $\chi_\Gamma$ is described as
\footnotesize
\begin{align}
\chi_\Gamma\left(T,B\right)&=\frac{1}{k_BT}\left\{\sum_{i}\rho_i\left|\Braket{i|\tilde{O}_\Gamma|i}\right|^2-\left(\sum_i\rho_i\Braket{i|\tilde{O}_\Gamma|i}\right)^2\right\}\nonumber\\
&+\sum_{j\neq i}\frac{\rho_j-\rho_i}{E_i-E_j}\left|\Braket{j|\tilde{O}_\Gamma|i}\right|^2. \tag{A$\cdot$11}
\end{align}
\normalsize
It can be seen that the diagonal terms of the matrix elements of the effective multipole $\tilde{O}_\Gamma$ correspond to the Curie terms, and the non-diagonal terms to the Van-Vleck terms from Eq. (A$\cdot$11). Here, $\rho_i=\frac{\rm{exp}\left(-\beta\it{E_i}\right)}{Z}$ means the probability based on statistical mechanics. The $\Gamma_3$-type electric quadrupole $O_v$ and electric hexadecapole $H_v$ used in the analysis of the present paper have matrix elements as follows. When $\lambda$ is not zero, the appropriate sum of the $9\times9$ matrices is considered as the matrix of the effective multipole $\tilde{O}_\Gamma$.

\begin{center}
\begin{eqnarray}
O_v&=&\bordermatrix{~ & \ket{\Gamma_3^{\left(\alpha\right)}} &  \ket{\Gamma_3^{\left(\beta\right)}}& \ket{\Gamma_4^{\left(\alpha\right)}}& \ket{\Gamma_4^{\left(\beta\right)}}& \ket{\Gamma_4^{\left(\gamma\right)}}& \ket{\Gamma_1}& \ket{\Gamma_5^{\left(\alpha\right)}}& \ket{\Gamma_5^{\left(\beta\right)}}& \ket{\Gamma_5^{\left(\gamma\right)}} \cr
              \bra{\Gamma_3^{\left(\alpha\right)}}  & \frac{8}{\sqrt{3}} & 0 &0&0&0&4\sqrt{\frac{35}{6}}&0&0&0\cr
              \bra{ \Gamma_3^{\left(\beta\right)} }& 0 & -\frac{8}{\sqrt{3}} &0&0&0&-4\sqrt{\frac{35}{6}}&0&0&0\cr
              \bra{\Gamma_4^{\left(\beta\right)}} & 0 & 0 &0&0&0&0&0&2\sqrt{7}&0\cr
              \bra{\Gamma_4^{\left(\gamma\right)}} & 0 & 0 &0&0&-14&0&-\sqrt{7}&0&0\cr
              \bra{\Gamma_1} & 4\sqrt{\frac{35}{6}} & -4\sqrt{\frac{35}{6}} &0&0&0&0&0&0&0\cr
              \bra{\Gamma_5^{\left(\alpha\right)}} & 0 & 0 &0&0&-\sqrt{7}&0&4&0&0\cr
              \bra{\Gamma_5^{\left(\beta\right)}} & 0 & 0 &0&2\sqrt{7}&0&0&0&0&0\cr
             \bra{ \Gamma_5^{\left(\gamma\right)}} &  0 & 0 &\sqrt{7}&0&0&0&0&0&-4 \cr}\nonumber\\
             \ \nonumber\           
\end{eqnarray}
\end{center}
\begin{center}
\begin{eqnarray}
H_v&=&\bordermatrix{~ & \ket{\Gamma_3^{\left(\alpha\right)}} &  \ket{\Gamma_3^{\left(\beta\right)}}& \ket{\Gamma_4^{\left(\alpha\right)}}& \ket{\Gamma_4^{\left(\beta\right)}}& \ket{\Gamma_4^{\left(\gamma\right)}}& \ket{\Gamma_1}& \ket{\Gamma_5^{\left(\alpha\right)}}& \ket{\Gamma_5^{\left(\beta\right)}}& \ket{\Gamma_5^{\left(\gamma\right)}} \cr
              \bra{\Gamma_3^{\left(\alpha\right)}}  & -160\sqrt{3} & 0 &0&0&0&12\sqrt{\frac{35}{6}}&0&0&0\cr
              \bra{ \Gamma_3^{\left(\beta\right)} }& 0 & 160\sqrt{3} &0&0&0&-12\sqrt{\frac{35}{6}}&0&0&0\cr
              \bra{\Gamma_4^{\left(\alpha\right)}} & 0 & 0 &-105&&0&0&0&0&45\sqrt{7}\cr
              \bra{\Gamma_4^{\left(\beta\right)}} & 0 & 0 &0&0&0&0&0&90\sqrt{7}&0\cr
              \bra{\Gamma_4^{\left(\gamma\right)}} & 0 & 0 &0&0&105&0&-45\sqrt{7}&0&0\cr
              \bra{\Gamma_1} & 12\sqrt{\frac{35}{6}} & -12\sqrt{\frac{35}{6}} &0&0&0&0&0&0&0\cr
              \bra{\Gamma_5^{\left(\alpha\right)}} & 0 & 0 &0&0&-45\sqrt{7}&0&75&0&0\cr
              \bra{\Gamma_5^{\left(\beta\right)}} & 0 & 0 &0&90\sqrt{7}&0&0&0&0&0\cr
             \bra{ \Gamma_5^{\left(\gamma\right)}} &  0 & 0 &45\sqrt{7}&0&0&0&0&0&-75\cr}\nonumber
\end{eqnarray}
\end{center}

\end{appendix}
\clearpage

\providecommand{\noopsort}[1]{}\providecommand{\singleletter}[1]{#1}%


\begin{thebibliography}{99}
\bibitem{enviroment}  The data with 110 MHz was obtained by connecting data from different refrigerators and different measurement environments. The data with 201 and 64 MHz were obtained by using a ${\rm^{3}He}$-${\rm^{4}He}$ dilution refrigerator and PPMS, respectively.
\end{thebibliography}
\end{document}